# Kasabach-Merritt syndrome arising from an Enteroatmospheric Fistula

Kunal Shah [1], Krishan Rajaratnam [2]

[1] University of Central Florida College of Medicine, Orlando, Florida

[2] New York University, New York, New York

‡ To whom the correspondence should be addressed: Kunal Shah - kshahmd@knights.ucf.edu

**Abstract**:

Kasabach-Merritt syndrome (KMS) is a rare, life-threatening condition that is characterized by profound thrombocytopenia, hypofibrinogenemia, elevated partial thromboplastin time, and may also be associated with microangiopathic hemolytic anemia. It is well established that this phenomenon is notably associated with the vascular *tumors* kaposiform hemangioendothelioma and tufted angioma; however, recent literature has suggested its presence in the settings of various vascular *malformations* (i.e. without neoplastic proliferation of endothelial cells). This report focuses on a patient in the first year of life, who experienced a chronic, consumptive coagulopathy in the setting of a highly vascular enteroatmospheric fistula. Sharing many features with the aforementioned syndrome, this anomaly suggests a novel association of the Kasabach-Merritt phenomenon with a unique vascular malformation. Although potentially lethal, Kasabach-Merritt syndrome can resolve with appropriate diagnosis and management; uncovering new associations can help to improve recognition and treatment in future cases.

**Introduction/Background:**
In 1940, Kasabach and Meritt observed a pediatric patient exhibiting thrombocytopenic purpura superimposed on a vascular tumor. This aberrant coincident was accompanied by increased bleeding and coagulation times, and treated with frequent blood product transfusions and, definitively, with radiotherapy.

The pathophysiology of Kasabach-Merritt syndrome (KMS) is largely theorized to be secondary to the native endothelial architecture of the vascular tumors Kaposiform Hemangioendothelioma (KHE) and Tufted Angioma (TA). The aforementioned tumors produce a hypervascular environment, histologically shown to include aberrant angiogenesis and lymphangiogenesis, which can instigate vascular instability. These conditions can induce turbulent blood flow and endothelial shearing, which abets the sequestration of platelets and coagulation factors. Severe complications such as microangiopathic hemolytic anemia and significant bleeding can arise. Sustained thrombocytopenia and consumptive coagulopathy from this phenomenon is collectively known as KMS [Lewis 2022].

Since Kasabach and Merritt's initial observations of the syndrome, many researchers observed similar coagulopathies as a result of the sequestration of platelets and coagulation factors in patients with the vascular tumors KHE and TA. This has led some to associate KMS exclusively with those vascular tumors [Lewis 2022]. Recent research, however, has shown symptoms reminiscent of KMS presenting in patients with vascular *malformations* (i.e. without endothelial neoplasm as seen in vascular tumors). As such, some have expanded upon KMS to be more inclusive of the broad spectrum of *all* vascular anomalies that can induce blood-flow conditions similar to that of the tumors KHE and TA (e.g. hematomas and Rendu-Osler disease [Famularo 2018]).

KMS treatment is largely supportive and aimed at preventing life-threatening complications while addressing the underlying vascular anomaly. Transfusion of blood products is generally utilized to mitigate corresponding thrombocytopenias, anemias, and coagulation factor deficiencies. Definitive treatment is surgical removal of the underlying vascular anomaly if feasible, as established in the cases of KHE and TA [Lewis 2022] but not in that of complicated or disseminated diseases (i.e. Rendu-Osler disease).

Despite the phenomenon's ambiguity surrounding its associations in literature, KMS has indisputably been demonstrated across sexes and ethnicities. Although cases generally present in infancy, there has been a subset of cases reported in adult populations as well. Additionally, KMS has led to mortality rates of up to 30 percent; causes of death include but are not limited to severe hemorrhage, cardiac failure, and local invasion specifically in the cases of the associated vascular tumors [Lewis 2022].

The recent expansion by researchers to include vascular anomalies beyond KHE and TA among the associations of KMS has unlocked a new avenue for understanding the phenomenon. Recognizing new associations is of critical importance to improve the condition's respective diagnosis and management in emerging cases. This paper presents an atypical case that further expands our knowledge of KMS and its associations: the apparent manifestation of KMS in the setting of an enteroatmospheric fistula with unusual hypervascularity.

**Case**:

A 23-day-old newborn arrives at the neonatal intensive care unit (NICU) with symptoms of multisystem dysfunction and unremitting thrombocytopenia, anemia, and coagulopathy. He was born at 23-weeks gestation with a rapidly declining progression at an alternate facility, which prompted swift transfer to a specialized hospital.

Before arrival at the NICU, the newborn suffered from a multitude of complications related to prematurity. APGAR scores were 6 and 8 at 5 and 10 minutes respectively. Weight was at the 72nd percentile, length was at the 10th percentile, and head circumference was below 5th percentile; growth status was appropriate for gestational age. The patient's condition began to deteriorate with the development of intraventricular hemorrhage, neonatal respiratory distress syndrome (NRDS), methicillin-resistant staphylococcus aureus (MRSA) bacteremia, anemia, electrolyte abnormalities, along with many additional adversities. Management included but was not limited to respiratory support with ventilation, frequent blood product transfusions, and the following medications: furosemide, cefepime, clindamycin, zosyn, acetaminophen, esmolol, and caffeine in an attempt to stabilize the patient. While all complications were being managed appropriately (i.e. with systemic support), the focus of this report will pertain to the evolution of this patient's unremitting coagulopathy, anemia, and thrombocytopenia over the course of 6 months in the setting of aberrant gastrointestinal malformations.

On arrival at the NICU, the patient continued to exhibit symptoms of the aforementioned complications. Vital signs revealed labile blood pressures and intermittent tachycardia. Comprehensive physical examination revealed a firm, distended abdomen. Laboratory results confirmed hyponatremia at 130 mmol/L, hyperkalemia at 5.9 mmol/L, elevated blood urea nitrogen (BUN) at 47 mg/dL, elevated alkaline phosphatase (ALP) at 1349 U/L, low total protein and albumin levels at 3.2 g/L and 1.8 g/L, respectively. Complete blood count showed white blood cells within acceptable limits, low hemoglobin at 11.4 g/dL, low hematocrit at 31%, and low platelets at $46 \times 10^3$ per μL. No active bleeding was observed at this time.

Patient's platelet counts, hemoglobin, and hematocrit levels remained low despite numerous packed red blood cell (pRBC) and single donor platelet (SDP) transfusions. On the 40th day of life, it was evident that the patient sustained significantly elevated levels of immature platelets, which suggested that the coagulopathy was consumptive in nature. This was further corroborated by the patient's elevated prothrombin time (PT) at 13.7 seconds and international normalized ratio (INR) at 1.2. Partial thromboplastin time (PTT) and fibrinogen activity levels were within normal limits at this time.

Over the next six months, the patient continued to rely heavily on frequent red blood cell, platelet, and fresh frozen plasma (FFP) transfusions to maintain acceptable cell counts. The patient's condition was further exacerbated by numerous systemic issues, most notably a progressive gastrointestinal failure that prompted a series of exploratory laparotomies beginning on the 42nd day of life. An extensive list of surgeries is summarized in table 1.

| Days of Life | Description of Exploratory Laparotomy |
|---|---|
| Day 42 | First laparotomy: Bowel exposed and placed under wound vacuum. |
| Day 45 | Irrigation: Observed area of necrosis with perforation and bowel leakage. |
| Day 49 | Irrigation: Observed segments of necrotic bowel and stool in the wound vacuum. Discovered 4 enteroatmospheric fistulae within viable bowel. |
| Day 51 | Irrigation: Verified the stability of viable bowel and enteroatmospheric fistulae. |
| Day 53 | Irrigation: Verified the stability of viable bowel and enteroatmospheric fistulae. |
| Day 56 | Irrigation: Additional enteroatmospheric fistula observed, incrementing the running total of known fistulae to 5. |
| Day 58 | Irrigation: Additional enteroatmospheric fistula observed, incrementing the running total of known fistulae to 6. |
| Day 60 | Irrigation: Fistulae irrigated. Meconium-like stool removed from one of the fistula. |
| Day 63 | Irrigation: Mobilization of bowel and ileostomy creation. |
| Day 65 | Irrigation. |
| Day 67 | Irrigation and ileostomy stabilization. |
| Day 70 | Ileostomy stabilization. |
| Day 72 | Ileostomy stabilization and fistulae examination. |
| Day 77 | Observed two additional fistulae. |
| Day 156 | Resection of massive chronic fistulae (fig. 1) and re-anastomoses of segmented bowel. Patient exhibited thrombocytopenia, severe bleeding, and cardiac failure during the procedure. Abdomen was left open. |
| Day 160 | Exploratory laparotomy to assess viability of remaining bowel after resection. |
| Day 162 | Partial closure of abdomen. |

| Day 165 | Partial closure of abdomen. |
| Day 169 | Partial closure of abdomen. |
| Day 172 | Partial closure of abdomen. |
| Day 175 | Complete closure accomplished. |

**Table 1.** Descriptions of all exploratory laparotomies performed in the patient.

The first laparotomy on the patient's 42nd day of life revealed necrotizing enterocolitis as the origin of the patient's overt, worsening gastrointestinal disease despite prior negative imaging; the abdomen was left open for future procedures and exposed contents were placed under vacuum. During an abdominal irrigation procedure performed on the patient's 49th day of life, multiple enteroatmospheric fistulae were discovered within viable regions of the bowel; also present were necrotic segments complicated by perforation and leakage of stool in the peritoneum. Additional fistulae were discovered over subsequent procedures. Ileostomy was created on day 63 of life with plans to re-anastomose fragmented bowel once feasible.

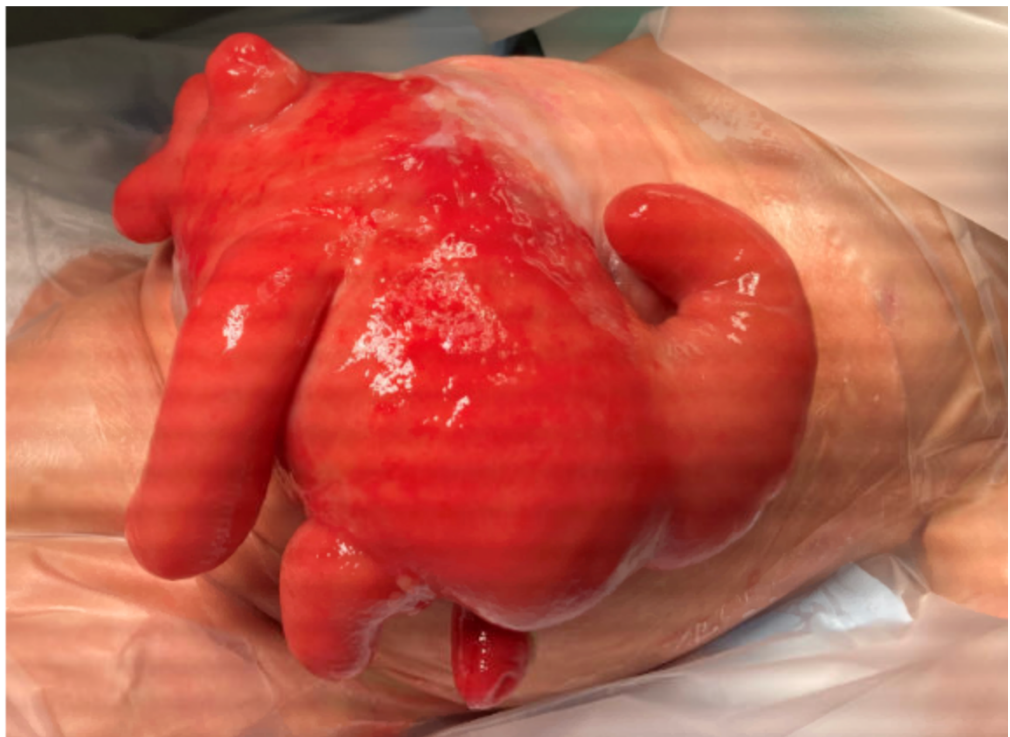

**Figure 1.** The hypervascular fistulous region of bowel resected from the patient.

Blood product transfusions at this time continued to be heavily relied upon and frequently administered to the patient at a rate of 40 per month to resolve the multifactorial cytopenias, especially due to the number of surgeries performed. The patient also exhibited bleeding from multiple mucosal and serosal sites throughout this time. On day 156, resection of the aberrant, fistulous region of bowel (fig. 1) was attempted and complicated by severe hemorrhage, bradycardia, and hypotensive shock in the operating room; this led to initiation of massive transfusion protocols and resuscitation of the patient via CPR with success. Majority of the patient's bleeding was observed to have occured from the bowel region in question, leading to the hypothesis that it was hypervascular in nature. This dissection was eventually completed, and a large, enteroatmospheric portion containing the fistulae was removed. Viable bowel segments were ultimately re-anastamosed. The abdomen remained open with plans to close gradually in a final series of laparotomies with concern for adequate healing of the abdominal contents and wall.

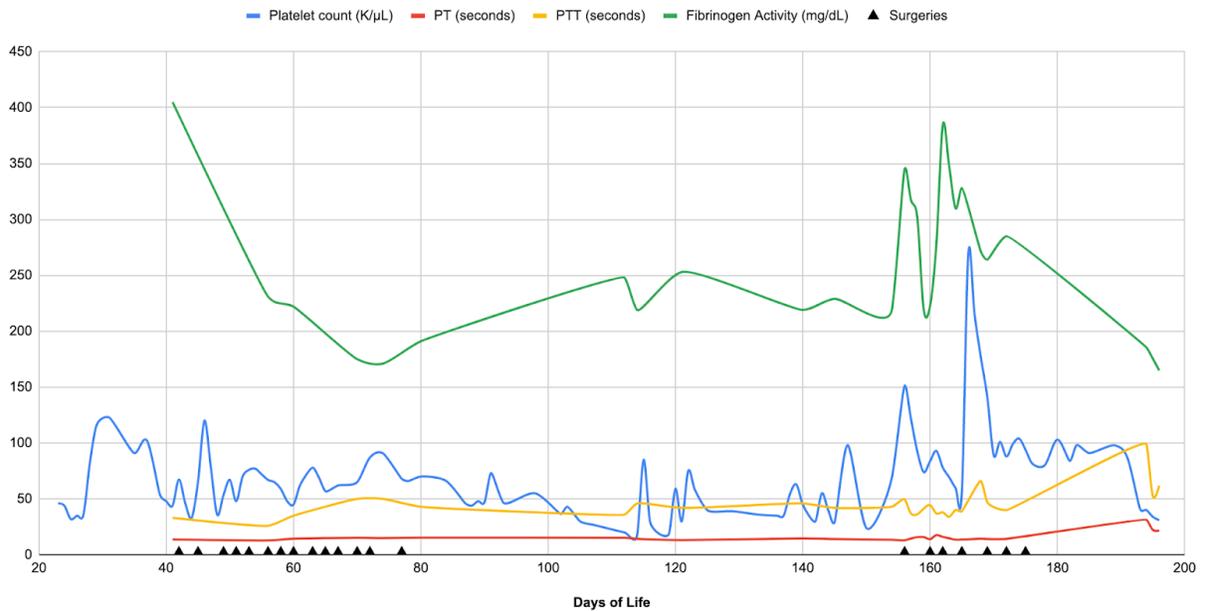

**Figure 2.** A line graph depicting platelet levels, PT, PTT, and fibrinogen activity of the patient while at the NICU.

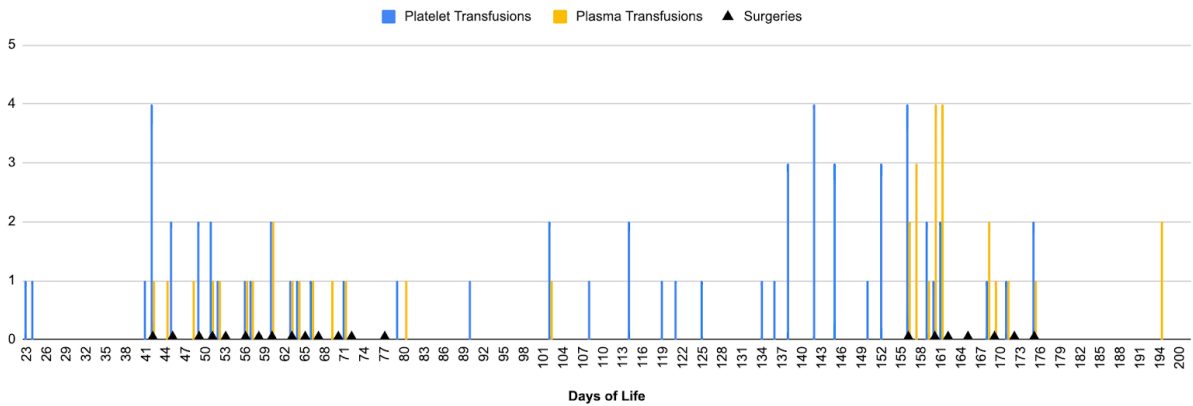

**Figure 3.** A chart depicting the number of platelet and plasma transfusions performed on the patient while at the NICU.

On day 162 of life, 6 days post fistulae resection, the patient was noted to have a significant improvement in the rate of platelets levels decline. As such, the frequency of respective transfusions declined as well. Although still low, platelet counts began to spontaneously maintain levels at ~90 × $10^3$/μL (fig. 2) without the need for *any* additional platelet transfusions beyond complete abdominal closure on day 175 of life (fig. 3).

PT and PTT initially showed evidence of similar stabilization with an average of ~14.3 seconds and ~40 seconds, respectively, after the aberrant bowel resection. This was further supported by a decreased frequency of plasma transfusions required by the patient allowing a cessation in plasma transfusions altogether after full closure on day 175, reminiscent of the course followed by platelet transfusions. The patient continued to exhibit no signs of obvious or internal bleeding.

On day 190, thrombocytopenia and coagulopathy returned in the setting of acute-on-chronic liver failure, as evident on figure 2. At this time, no platelet transfusions and 2 plasma transfusions were administered (fig 3.) to appease the diminishing hepatic synthetic function of coagulation factors.

**Discussion**:

This case examined an infant who was born prematurely with findings largely consistent with a Kasabach-Merritt phenomenon, manifesting in the context of a hyper-vascular, fistulous region of bowel. The patient had symptoms and laboratory values significant for unremitting thrombocytopenia, anemia, and coagulopathy throughout his admission, which dramatically improved after surgical resection of this bowel segment.

Frequent transfusions of red blood cells, platelets, and plasma were required throughout hospitalization. Frequent bleeding observed in both mucosal and serosal sites raised further concern for the patient's multisystem dysfunction; however the patient's profound thrombocytopenia with a significantly elevated immature fraction and abnormal coagulation studies pointed more towards a consumptive coagulopathy as the principal origin.

Multiple exploratory laparotomies were conducted due to the patient's worsening gastrointestinal dysfunction, and a large, fistulous region of bowel was discovered (fig. 1). It was noted to be extensively vascular in nature, as evidenced by the massive blood loss experienced by the patient during its dissection and resection on day 156 of life. The patient subsequently experienced profound bradycardia and hypotension, which necessitated successful resuscitation. Over the following 20 days the patient's platelet levels and coagulation studies began to stabilize spontaneously. This was evidenced by a dramatic slowing in the rate of platelet level decline and plateauing at an acceptable count. Coagulation studies showed similar patterns with their respective measures. Beyond day 175 of life (complete closure of the abdomen), the patient required no additional blood product transfusions in the setting of a removed enteroatmospheric fistula. No obvious bleeding from mucosal or serosal sites was observed.

The patient's significant improvement after resection of the hypervascular region of bowel was substantiated by stabilized cell counts and pertinent laboratory studies. Combined with the profound blood loss experienced during the procedure and a dramatic decrease in the frequency of blood product transfusions shortly after, the evidence indicates the existence of a local, consumptive coagulopathy with systemic effects originating from the region of bowel resected on day 156 of life. Therefore, this same evidence indicates the presence of a Kasabach-Merritt phenomenon arising from a vascular, fistulous region of bowel.

Considering that literature states an established connection between KMS and the vascular tumors KHE and TA [Lewis 2022], this phenomenon occurring as part of an enteroatmospheric, fistulous region of bowel is uncommon and has not been reported before.

**Conclusion**

Kasabach-Merritt syndrome is a rare, life-threatening coagulopathy that is characterized by persistent thrombocytopenia, anemia, and abnormal coagulation studies; more severe complications can include microangiopathic hemolytic anemia and massive bleeding. The phenomenon is believed to be a product of platelet and coagulation factor sequestration and activation in the setting of vascular tumors. Of note, KMS has been almost exclusively established in the vascular tumors kaposiform hemangioendothelioma and tufted angioma, which typically occur in infancy.

Recent literature suggests that the phenomenon may occur in a number of vascular *malformations* (i.e. non-neoplastic) as well; this is due in part to the shared ability of many vascular anomalies to form aberrant architecture and endothelial structures similar to those seen regularly in vascular tumors. This report is one of many in support of KMS arising from atypical vascular anomalies. As such, evidence calls for the definition of Kasabach-Merritt syndrome to be more inclusive of *all* vascular anomalies with regards to the origins of emerging cases.

**References:**

1. Lewis D, Vaidya R. Kasabach Merritt Syndrome. 2022 Jan 19. In: StatPearls [Internet]. Treasure Island (FL): StatPearls Publishing; 2022 Jan–. PMID: 30085595.
2. Famularo G. Kasabach-Merritt Syndrome. Am J Med. 2020 Dec;133(12):e747. doi: 10.1016/j.amjmed.2020.02.032. PMID: 33248667.
3. Mahajan P, Margolin J, Iacobas I. Kasabach-Merritt Phenomenon: Classic Presentation and Management Options. Clin Med Insights Blood Disord. 2017 Mar 16;10:1179545X17699849. doi: 10.1177/1179545X17699849. PMID: 28579853; PMCID: PMC5428202.
4. Borst, Alexandra J., and Taizo A. Nakano. "Targeting inflammation-induced Kasabach-Merritt phenomenon." Blood, The Journal of the American Society of Hematology 139, no. 11 (2022): 1603-1605.